\documentclass[twocolumn]{aastex63}
\usepackage{ulem}
\usepackage{amsmath}
\graphicspath{{./fig/}}

\shorttitle{Super-Earths and their atmospheres}
\shortauthors{Ogihara and Hori}

\begin{document}

\title{Unified simulations of planetary formation and atmospheric evolution: Effects of pebble accretion, giant impacts, and stellar irradiations on super-Earth formation}

\correspondingauthor{Masahiro Ogihara}
\email{masahiro.ogihara@nao.ac.jp}

\author[0000-0002-8300-7990]{Masahiro Ogihara}
\affiliation{National Astronomical Observatory of Japan,
2-21-1, Osawa, Mitaka,
181-8588 Tokyo, Japan}

\author[0000-0003-4676-0251]{Yasunori Hori}
\affiliation{National Astronomical Observatory of Japan,
2-21-1, Osawa, Mitaka,
181-8588 Tokyo, Japan}
\affiliation{Astrobiology Center,
2-21-1, Osawa, Mitaka,
181-8588 Tokyo, Japan}

\begin{abstract}

A substantial number of super-Earths have been discovered, and atmospheres of transiting super-Earths have also been observed by transmission spectroscopy. Several lines of observational evidence indicate that most super-Earths do not possess massive H$_2$/He atmospheres. However, accretion and retention of less massive atmospheres on super-Earths challenge planet formation theory.
We consider the following three mechanisms: (i) envelope heating by pebble accretion, (ii) mass loss during giant impacts, and (iii) atmospheric loss by stellar X-ray and EUV photoevaporation. We investigate whether these mechanisms influence the amount of the atmospheres that form around super-Earths.
We develop a code combining an \textit{N}-body simulation of pebble-driven planetary formation and an atmospheric evolution simulation.
We demonstrate that the observed orbital properties of super-Earths are well reproduced by the results of our simulations. However, (i) heating by pebble accretion ceases prior to disk dispersal, (ii) the frequency of giant impact events is too low to sculpt massive atmospheres, and (iii) many super-Earths having H$_2$/He atmospheres of $\gtrsim 10$\,wt\% survive against stellar irradiations for 1\,Gyr.
Therefore, it is likely that other mechanisms such as suppression of gas accretion are required to explain less massive atmospheres ($\lesssim 10$\,wt\%) of super-Earths.
\end{abstract}


\section{Introduction} \label{sec:intro}

A significant number of close-in planets have been discovered primarily via transit photometry and radial velocity surveys, which have revealed a variety of the characteristics of these planets. 
The orbital properties of super-Earths are of particular importance for unveiling their formation histories. 
Period ratio distributions of adjacent planets in multiple super-Earth systems show that although some planet pairs are in near mean motion resonances \citep{2014ApJ...790..146F}, the majority of super-Earths are not in mean motion resonances. Formation of super-Earths has been investigated by \textit{N}-body simulations of planet formation \citep[e.g.][]{2007ApJ...654.1110T, 2009ApJ...699..824O, 2014A&A...569A..56C}.
Several previous studies have pointed out an important issue in reproducing the orbital properties of super-Earths, in which super-Earths pile up at the inner edge of the disk due to rapid type I migration \citep[e.g.,][]{2015A&A...578A..36O, 2017A&A...607A..67M}. In recent studies, several authors succeeded in reproducing most of the observed characteristics using a disk evolution model that includes disk winds \citep{2018A&A...615A..63O} or pebble-driven planet formation model \citep{2019A&A...627A..83L, 2019arXiv190208772I}.

Recently, atmospheres of transiting super-Earths have been observed by transmission spectroscopy using ground-based telescopes and Hubble space telescope and {\it Spizter}. Transmission spectra in the atmospheres of super-Earths show no prominent absorption features, which indicate either a hydrogen-poor atmosphere or clouds/haze in a hydrogen-rich one \citep[e.g.][]{2010Natur.468..669B,2014Natur.505...66K}. Masses and radii of exoplanets also reveal the existence of super-Earths having atmospheres, e.g., Kepler-11 planets. Interior modeling of observed super-Earths suggests that most of them would possess less massive ($\sim 0.1-10\%$ by mass) H$_2$/He atmospheres \citep[e.g.][]{2014ApJ...792....1L}. 

Accretion and retention of less massive atmospheres of super-Earths are puzzling problems for their origin. \textit{Kepler} planets are clustered around a core mass of $M_{\rm core} = 3-5 \,M_\oplus$ \citep{2017ApJ...847...29O, 2019ApJ...878...36L}. Such massive cores undergo runaway gas accretion within the disk lifetime and accumulate massive H$_2$/He atmospheres from the protoplanetary disk \citep[e.g.][]{2012ApJ...753...66I,2014ApJ...797...95L}. In addition, close-in super-Earths ($r < 1 {\rm \,au}$) are over 10 times as common as close-in giant planets \citep[e.g.][]{2010Sci...330..653H, 2013ApJ...766...81F}. These imply that  super-Earth cores are likely to avoid accreting massive atmospheres from the disk in the formation stage.
Several possible solutions have been proposed to explain the origin of super-Earths with less massive atmospheres: high opacities (or metallicities) in the planetary envelope \citep[e.g.][]{2014ApJ...797...95L}, the continual recycling of the accreting gas around the planet \citep[e.g.][]{2015MNRAS.447.3512O, 2017MNRAS.471.4662C, 2018MNRAS.479..635K, 2019A&A...623A.179K}, the delay of gas accretion by polluted envelopes \citep{2020A&A...634A..15B}, and disk dissipation \citep[][]{2012ApJ...753...66I,2020ApJ...889...77H}. We have recently proposed the possibility of suppressing envelope accretion via a limit due to the disk accretion rate \citep{2018ApJ...867..127O}, and similar solutions have been discussed in other studies \citep[e.g.][]{2019ApJ...878...36L, 2019MNRAS.487..681G}.

The atmospheres of super-Earths likely originate from accretion of the disk gas in the formation stage, if they retain cloudy hydrogen-rich atmospheres as suggested by their featureless transmission spectra. In this paper, we develop a unified numerical model of planetary formation and atmospheric evolution to investigate whether super-Earths with less massive atmospheres can form. We consider the following three possible mechanisms: envelope heating by pebble accretion, mass loss during giant impacts, and atmospheric loss by stellar X-ray and EUV photoevaporation. 
The heating of the envelope by pebble accretion inhibits the accumulation of massive atmospheres \citep[e.g.][]{2014A&A...572A..35L}. We determine the onset of runaway gas accretion onto the core, namely, the critical core mass, by calculating interior structures of planets under pebble accretion. The envelope of planets can be eroded through giant impacts triggered by dynamical instability in the late stage of planet formation. We implement the effect of impact erosion into the \textit{N}-body simulation using an empirical formula for mass loss during giant impacts. We also consider atmospheric loss from planets by stellar X-ray and EUV (XUV) irradiations after disk dissipation \citep[e.g.][]{2017ApJ...847...29O}, which are closely related to the radius valley in the radius-period distribution of small planets \citep{2017AJ....154..109F}.

Atmospheric accretion onto super-Earths depends on formation history and disk evolution; therefore, it is crucial to self-consistently follow the planetary growth, the orbital evolution, and the atmospheric evolution.  
For the disk evolution model, we do not use a simple power-law disk model such as the minimum mass solar nebula. Instead, we use the result of a one-dimensional disk evolution simulation that takes into account the effects of magnetically driven disk winds \citep{2016A&A...596A..74S}.
Our primary aim is to examine whether the above mechanisms are able to address the issue of the formation of super-Earths with less massive atmospheres. We also examine whether the observed orbital properties (e.g., the period ratio of adjacent planets) can be reproduced by the results of the new simulations taking into account pebble accretion and atmospheric evolution.

This paper is structured as follows. In Section~\ref{sec:model}, we describe our model of a unified simulation of planetary formation and atmospheric evolution: planet formation by pebble accretion in an evolving disk and atmospheric loss by giant impacts and stellar XUV irradiations. 
In Section~\ref{sec:n-body}, we present the results of {\it N}-body simulations of super-Earth formation for up to 50\,Myr. In Section~\ref{sec:PE}, we present the results of long-term simulations of atmospheric loss from planets by photoevaporation for 1\,Gyr. In Section~\ref{sec:disk_limit}, we also present the results of simulations that include a suppression of gas accretion onto planetary cores, i.e., a limit on the atmospheric accretion by disk accretion. In Section~\ref{sec:conc}, we present discussion and our conclusions.

\section{Model} \label{sec:model}

Our study consists of two simulations: the first is a unified simulation of an \textit{N}-body simulation of planetary formation and the atmospheric evolution, which is simulated for 50\,Myr, and the second is a long-term (1\,Gyr)  atmospheric loss simulation of simulated planets following the formation stage.

\subsection{Unified simulations of formation and atmospheric evolution}

\subsubsection{Pebble accretion}\label{sec:pebble}
In this paper, we use simple prescriptions for the pebble accretion. The pebble accretion rate for the two-dimensional (2D) accretion mode is \citep[e.g.][]{2015Icar..258..418M}
\begin{equation}
\dot{M}_{\rm 2D} = 2 r_{\rm eff} v_{\rm acc} \Sigma_{\rm pb},
\end{equation}
where $r_{\rm eff}, v_{\rm acc}$, and $\Sigma_{\rm pb}$ are the effective radius for pebble accretion, the accretion velocity of the pebble onto the planetary core, and the surface density of pebbles in the disk, respectively. The 2D accretion mode means that the pebble scale height $H_{\rm pb}$ is smaller than the effective pebble cross section $r_{\rm eff}$. The pebble scale height is related to the gas scale height $H$ such that \citep{2007Icar..192..588Y}
\begin{equation}\label{eq:hpb}
H_{\rm pb} \sim \sqrt{\frac{\alpha}{\alpha + \tau_{\rm s}}} H \sim \sqrt{\frac{\alpha}{\tau_{\rm s}}} H,
\end{equation}
where $\alpha$ is the turbulent diffusion parameter in the $\alpha$-viscosity prescription \citep{1973A&A....24..337S} and $\tau_{\rm s}$ is the Stokes number. In the three-dimensional (3D) accretion mode, the pebble accretion rate is
\begin{equation}
\dot{M}_{\rm 3D} = \pi r_{\rm eff}^2 v_{\rm acc} \rho_{\rm pb},
\end{equation}
where $\rho_{\rm pb}$ is the density of pebbles on the midplane. The effective radius for pebble accretion is expressed as
\begin{equation}\label{eq:r_eff}
r_{\rm eff} = \left(\frac{\tau_{\rm s}}{0.1}\right)^{1/3} R_{\rm GP},
\end{equation}
where $R_{\rm GP}= \min(R_{\rm B}, R_{\rm H})$ is the effective radius for the gravitational pull \citep{2012A&A...544A..32L, 2015Icar..258..418M}. The Bondi radius is expressed using the relative velocity between the pebble and the core such that $R_{\rm B} = GM/ \Delta v^2$. The Hill radius is $R_{\rm H} = a [M/(3M_*)]^{1/3}$, where $a$ is the semi-major axis of the accreting core. We introduce a reduction factor of $r_{\rm eff}$ when the stopping time of pebbles due to gas drag is longer than the Keplerian frequency, i.e., $\tau_{\rm s} > 0.1$ \citep{2010A&A...520A..43O, 2012ApJ...747..115O, 2016A&A...591A..72I},
\begin{eqnarray}
r_{\rm eff}|_{\tau_{\rm s} > 0.1} &&= r_{\rm eff} \nonumber  \\
\times && \exp\left( - \left\{ \frac{\tau_{\rm s}}{\min(2,4 [M/M_*]/ \eta^3))}\right\}^{0.65}\right),
\end{eqnarray}
where $M$ is the planetary mass, $M_*$ is the mass of the central star, and $\eta\,(= -1/2 (H/r)^2 \partial \ln P / \partial \ln r)$ represents the deviation from the Keplerian motion of the disk gas due to the radial pressure gradient.

The accretion velocity of the pebble onto the accreting core is defined as $v_{\rm acc} = \Delta v + r_{\rm eff} \Omega_{\rm K}$, where $\Omega_{\rm K}$ is the Keplerian frequency. In the Bondi regime, $v_{\rm acc} \simeq \Delta v$, and $v_{\rm acc} \simeq r_{\rm eff} \Omega_{\rm K}$ for the Hill regime.
Here,  the relative velocity $\Delta v$ is given by
\begin{equation}
\Delta v \simeq   \frac{\sqrt{4 \tau_{\rm s}^2 + 1}}{\tau_{\rm s}^2 +1} \eta v_{\rm K} \simeq \eta v_{\rm K},
\end{equation}
where $v_{\rm K}$ is the Keplerian velocity\footnote{For simplicity, we do not take into account the dependence of the pebble accretion rate on the eccentricity and the inclination.}.

The pebble accretion rate also depends on the pebble surface density. In this study, a steady-state pebble surface density is calculated such that
\begin{eqnarray}
\Sigma_{\rm pb} = \frac{\dot{M}_{\rm pb}}{2 \pi r v_r},
\end{eqnarray}\\
where $v_{\rm r}$ is the radial drift velocity \citep{1977MNRAS.180...57W,1986Icar...67..375N},
\begin{equation}
v_r = -\frac{2\tau_{\rm s}}{\tau_{\rm s}^2 +1} \eta v_{\rm K}.
\end{equation}
There is a huge uncertainty in the pebble mass flux which strongly depends on properties and time evolution of the protopalnetary disk; therefore, we treat the pebble mass flux, $\dot{M}_{\rm pb}$, as a parameter in the same way as in previous studies \citep[e.g.,][]{2019A&A...627A..83L, 2019arXiv190208772I, 2019A&A...623A..88B}. An exponential decay is assumed for the pebble mass flux on a timescale of 1\,Myr\footnote{Although we adopt the decay timescale of 1\,Myr as in \cite{2019A&A...627A..83L}, our results in this paper do not depend sensitively on the decay timescale. This is because, as seen in Section~\ref{sec:suppress}, after planets reach 1 Earth mass, they can grow to the pebble isolation mass in a short time. The time to reach the pebble isolation mass is usually much shorter than the decay timescale.}. Regarding the pebble size, we consider 1-mm-size silicate pebbles, which are produced by either dust coagulation inside the snow line or by sublimation of icy pebble near the snow line. The evolution of the pebble size and the Stokes number are very uncertain; therefore, we fix the pebble size as in previous studies \citep[e.g.,][]{2015Icar..258..418M, 2019arXiv190208772I}. 
We also consider a filtering of pebbles \citep[e.g.][]{2014A&A...572A.107L, 2014A&A...572A..72G}. The pebble accretion rate is reduced in accordance with the amount of pebbles accreted on outer planets.

When the mass of a planetary core exceeds the pebble isolation mass, the pebble accretion ceases \citep[e.g.][]{2012A&A...546A..18M}. We use a revised formula of the pebble isolation mass \citep{2018A&A...612A..30B}:
\begin{eqnarray}\label{eq:miso}
M_{\rm iso,pb} &=& 25 \left( \frac{H/r}{0.05}\right)^3 \left\{ 0.34 \left( \frac{-3}{\log{10}(\alpha)}\right)^4 + 0.66 \right\} \nonumber \\
&& \times \left( 1 - \frac{ \frac{\partial \ln P}{\partial \ln r}+ 2.5}{6}\right) M_\oplus.
\end{eqnarray}
When a planet exceeds the pebble isolation mass, pebble accretion onto the inner planets is also quenched in our simulations.

\subsubsection{Initial condition}
Similar to previous studies \citep{2017A&A...607A..67M, 2019A&A...627A..83L}, we start simulations with Lunar-mass embryos ($M = 0.01 M_\oplus$). The transition mass from the Bondi regime to the Hill regime is given by \citet{2012A&A...544A..32L}
\begin{equation}
M_{\rm t} = \sqrt{\frac{1}{3}} \frac{\Delta v^3}{G \Omega_{\rm K}},
\end{equation}
which is basically smaller than $0.01 M_\oplus$. Therefore, we focus on the Hill regime in this paper. 
Note that the pebble accretion rate for the Bondi regime is orders of magnitude smaller than that for the Hill regime \citep[e.g.][]{2017AREPS..45..359J}. Thus, it takes long time to grow to Lunar-mass embryos only by pebble accretion.
Embryos are initially placed between $r = 0.1 {\rm \,au}$ and $2 {\rm \,au}$ with a logarithmic spacing. The total initial mass of embryos is $1 \, M_\oplus$.

\subsubsection{Disk evolution}\label{sec:disk}
We use a disk model developed by \cite{2016A&A...596A..74S}, as in previous studies \citep[see also][]{2018A&A...615A..63O, 2018A&A...612L...5O}. In this model, the disk evolves via viscous accretion, the accretion driven by the wind torque (wind-driven accretion), and the mass loss due to disk winds. To obtain the long-term evolution, we numerically solve the following diffusion equation:
\begin{eqnarray}\label{eq:diffusion}
\frac{\partial \Sigma_{\rm g}}{\partial t} &=& \frac{1}{r} \frac{\partial}{\partial r} \left[\frac{2}{r\Omega} \left\{ \frac{\partial}{\partial r} (r^2 \Sigma_{\rm g} \alpha_{r,\phi} c_{\rm s}^2) + r^2 \alpha_{\phi,z}\frac{\Sigma_{\rm g} H \Omega^2}{2 \sqrt{\pi}} \right\} \right]\nonumber \\
 &&- C_{\rm w} \frac{\Sigma_{\rm g} \Omega}{\sqrt{2 \pi}},
 \label{eq:wind}
\end{eqnarray}
where $\Sigma_{\rm g}$ is the disk surface density. The first term on the right-hand side of Eq.\,(\ref{eq:wind}) is the viscous accretion, the second term is the wind-driven accretion, and the third term is the mass loss due to the disk winds. 
As stated in previous studies, the second term dominates the evolution of the gas surface density. The parameter for the turbulent viscosity $\alpha_{r,\phi}$ is set to $8 \times 10^{-3}$. 
The values of $\alpha$ in Equations.(\ref{eq:hpb}) and (\ref{eq:miso}) are set to $\alpha_{r,\phi}$.
In the previous study of \cite{2018A&A...615A..63O}, formation of super-Earths was investigated for various disk parameters. They found that observed orbital properties of super-Earths (e.g., period ratio) can be well reproduced in cases with $\alpha_{r,\phi} = 8 \times 10^{-3}$. We also use the same value of alpha viscosity. A discussion of the dependence on the viscosity is presented in Section~\ref{sec:conc}.
The parameter for the wind-driven accretion $\alpha_{\phi,z}$ \citep[see Eq.\,(30) in][]{2016A&A...596A..74S} increases with decreasing the plasma beta \citep[e.g.][]{2013ApJ...772...96B}. The parameter for the wind mass loss $C_{\rm w}$ is set to $2 \times 10^{-5}$.
Although the stellar XUV-induced photoevaporation would result in heating the gas at the disk surface which drives mass loss from the disk \citep[e.g.][]{2020MNRAS.492.3849K}, the photoevaporative effect on the disk is not taken into account in this paper. The effect of disk photoevaporation will be investigated in our next paper.

The global evolution of the gas surface density is significantly different from the minimum-mass solar nebula model \citep{1981PThPS..70...35H}; for example, the gas surface density decreases in the inner region $(r \lesssim 1 {\rm \,au})$ \citep[see Figure~1(a) in][]{2018A&A...615A..63O}. 
Such a decrease in the gas surface density in the close-in region is also seen in several analytical models of disk evolution that include effects of disk winds \citep[e.g.][]{2018MNRAS.475.5059K,2019ApJ...879...98C}. Note that \cite{2016ApJ...821...80B} derived a different disk evolution model from the one-dimensional disk evolution simulation, including disk winds. In their model, the gas surface density is not depleted in the close-in region. Implications of this different disk profile for orbital evolution are discussed in Section~\ref{sec:conc}.
Thermal evolution of the disk is determined by the viscous heating and the radiative equilibrium (see Section 2.4 in \citealt{2016A&A...596A..74S} for details).

\subsubsection{Envelope accretion}\label{sec:model_env}
 
Planets accrete H$_2$/He envelopes from the protoplanetary disk. 
The growth of a planetary envelope proceeds slowly while the concurrent accretion of gas and solid material occurs.
Once a planetary core grows to the critical core mass, its envelope, which can no longer maintain hydrostatic equilibrium, enters the runaway gas accretion phase. The critical core mass increases with increasing atmospheric heating due to the accretion of solids such as pebbles and planetesimals.
Assuming a pebble accretion rate ranging from $10^{-12}\,M_\oplus$\,yr$^{-1}$ to $10^{-2}\,M_\oplus$\,yr$^{-1}$,
we repeatedly calculate the 1D hydrostatic structure of a planet in the same way as \citet{2010ApJ...714.1343H}.
For simplicity, incoming pebbles are assumed to contribute little to the enhancement of the grain and gas opacities in the envelope.
We obtained the following formula for the critical core mass as a function of the pebble accretion rate (see Appendix~\ref{app:mcrit}):
\begin{eqnarray}\label{eq:mcrit}
M_{\rm crit} = 13 \left(\frac{\dot{M}_{\rm pb}}{10^{-6} \,M_\oplus \, {\rm yr}^{-1}}\right)^{0.23} M_\oplus.
\end{eqnarray}
In our \textit{N}-body simulations,
we assume that gas accretion onto planets starts when they reach the critical core mass\footnote{Planets with cores smaller than the critical core mass can accrete a small amount of H$_2$/He envelope from the disk. However, this does not affect our conclusion.}.

The envelope accretion rate is given by
\begin{eqnarray}\label{eq:mdot_min}
\dot{M}_{\rm env} = \min(\dot{M}_{\rm KH}, \dot{M}_{\rm hydro}, \dot{M}_{\rm disk}), 
\end{eqnarray}
where $\dot{M}_{\rm KH}$ is the gas accretion rate determined by the gravitational contraction of the planetary envelope,
$\dot{M}_{\rm hydro}$ is the gas capture rate derived from the hydrodynamics of the gas flow around the planet,
and $\dot{M}_{\rm disk}$ is the supply limit of the disk gas.
We assume that the planets accrete the local gas while conserving angular momentum \citep[e.g.,][]{2014ApJ...797....1K}.
The gas accretion rate $\dot{M}_{\rm KH}$ during Kelvin--Helmholtz contraction \citep{2010ApJ...714.1343H} is described as 
\begin{equation}
\dot{M}_{\rm KH} = 10^{-8} \left(\frac{M_{\rm core}}{M_\oplus}\right)^{3.5} M_\oplus \,{\rm yr}^{-1}.
\end{equation}
We adopt the gas capture rate $\dot{M}_{\rm hydro}$ given in \citet{2016ApJ...823...48T}:
\begin{eqnarray}
\dot{M}_{\rm hydro} &=& 0.29 \left(\frac{H}{r}\right)^{-2} \left(\frac{M}{M_*}\right)^{4/3} r^2 \Omega_{\rm K} \Sigma_{\rm min}\\
&=& 0.29 \left(\frac{H}{r}\right)^{-2} \left(\frac{M}{M_*}\right)^{4/3} r^2 \Omega_{\rm K} \frac{\Sigma_{\rm g}}{1 + 0.04 K}\\
K &=&\left( \frac{M}{M_*} \right)^2 \left( \frac{H}{r} \right)^{-5} \alpha^{-1}_{r,\phi},\label{eq:TT12}
\end{eqnarray}
where the gas surface density in the gap, $\Sigma_{\rm min}$, is expressed using a parameter $K$ \citep{2015MNRAS.448..994K}. Then, the gas supply rate throughout the global disk accretion is calculated such that
\begin{equation}\label{eq:mdot_disk}
\dot{M}_{\rm disk} = \max(\dot{M}_{\rm visc}, \dot{M}_{\rm wind}),
\end{equation}
where $\dot{M}_{\rm visc}$ and $\dot{M}_{\rm wind}$ indicate the viscous accretion and the wind-driven accretion, respectively. The formulas for these accretions are given by \citet{2016A&A...596A..74S}
\begin{eqnarray}
\dot{M}_{\rm visc} = \frac{2 \pi}{\Omega} \alpha_{r,\phi} c^2_{\rm s} \Sigma_{\rm g},\\
\dot{M}_{\rm wind} = 2 \sqrt{2 \pi} \alpha_{\phi,z} r c_{\rm s} \Sigma_{\rm g}.
\end{eqnarray}
Recent magnetohydrodynamic simulations suggest that wind-driven accretion may dominate over viscous accretion ($\dot{M}_{\rm visc} < \dot{M}_{\rm wind}$) in the inner disk (see also Figs.~3 and 9 in \citet{2018ApJ...867..127O}).
Provided that the wind-driven accretion does not contribute to the envelope accretion, namely, $\dot{M}_{\rm disk} = \dot{M}_{\rm visc}$,
massive planets such as super-Earths can avoid runaway gas accretion \citep{2018ApJ...867..127O}.
We also consider a limit on the disk accretion for the \textit{N}-body simulations shown in Section~\ref{sec:disk_limit} as in \citet{2018ApJ...867..127O}.

\subsubsection{Type I/II migration}

Planets with masses larger than $\sim 0.1\,M_\oplus$ undergo type I migration. The type I migration torque is determined by the superposition of the Lindblad torque and the corotation torque, which depend on the properties of the gas disk. In particular, the corotation torque can be positive when the surface density slope is positive, leading to a suppression of type I migration \citep[e.g.][]{2015A&A...584L...1O}. In this study, we use prescriptions that include the saturation of the corotation torque. For a detailed description, the reader is referred to Eqs.~(50)-(53) in \cite{2011MNRAS.410..293P}.
The damping of the eccentricity and inclination by density waves is also considered in this study. As in previous studies \citep[e.g.,][]{2019A&A...627A..83L}, the actual formulas for the damping forces are based on \cite{2008A&A...482..677C}, in which a reduction of the damping is included for planets in eccentric and inclined orbits \citep[e.g.,][]{2014MNRAS.445..479C}.

Larger planets start to carve a density gap in the protoplanetary disk, and the migration mode shifts from type I to type II. The migration timescale for type II migration and the transition phase are expressed using the $K$ coefficient in \cite{2018ApJ...861..140K} such that 
\begin{eqnarray}
t_{a, {\rm II}} = \frac{\Sigma_{\rm g}}{\Sigma_{\rm min}} t_{a, {\rm I}} \simeq (1+0.04K) t_{a, {\rm I}},
\end{eqnarray}
where $t_{a, {\rm I}}$ is the type I migration timescale and $K$ is equivalent to Eq.~(\ref{eq:TT12})\footnote{We assume that the parameter $K$ is determined by the turbulent $\alpha_{r,\phi}$. This assumption also used in \cite{2018ApJ...864...77I}, in which $\alpha_{r,\phi}$ and $\alpha_{\phi,z}$ correspond to $\alpha_{\rm vis}$ and $\alpha_{\rm acc}$, respectively. The wind-driven parameter $\alpha_{\phi,z}$ may also contribute to $K$. This should be investigated using hydrodynamical simulations.}. Note that \cite{2018ApJ...861..140K} obtained the above formula using the results of hydrodynamical simulations under the isothermal approximation. It is likely that this treatment is also valid for non-isothermal cases (see also Section 6 in \citealt{2018ApJ...861..140K}).
Even though the eccentricity damping timescale for massive planets is not well understood, we also multiply the eccentricity damping timescale by a factor of $(1 + 0.04 K)$.

\subsubsection{Atmospheric loss by giant impacts}

When planets collide, 
the global ground motion induced by shock waves can mechanically blow off atmospheres
\citep[e.g.][]{2003Icar..164..149G,2015Icar..247...81S}. We use a scaling law obtained from 3D hydrodynamic simulations of giant impacts \citep{2014LPI....45.2869S} for the blow off of the atmosphere of a planet during a giant impact.
\citet{2014LPI....45.2869S} defined a specific impact energy $Q_{\rm S}$ for a collision
between planets with core masses of $M_1$ and $M_2$ such that
\begin{eqnarray}\label{eq:Qs}
Q_{\rm S} = Q_{\rm R} \left( 1 + M_2/M_1\right) (1-b),
\end{eqnarray}
where
\begin{eqnarray}
Q_{\rm R} = \frac{1}{2} \frac{M_1 M_2}{(M_1 + M_2)^2} v_{\rm imp}^2
\end{eqnarray}
is used from \citet{2012ApJ...745...79L}
and $v_{\rm imp}$ is the impact velocity. The impact parameter is denoted by $b$, where $b = 0$ means a head-on collision. Then, the atmospheric loss fraction from the planet with $M_{\rm core}=M_1$ is approximately given by
\begin{eqnarray}
 L = \left\{ \begin{array}{ll}
    1 & (10^{7.76} < Q_{\rm S} ) \\
    0.562 \log_{10}{Q_{\rm S}} - 3.37& (10^{6.35} < Q_{\rm S} \leq 10^{7.76})\\
    0.0850 \log_{10}{Q_{\rm S}} - 0.340& (Q_{\rm S} \leq 10^{6.35}).
  \end{array} \right.
\end{eqnarray}

\subsubsection{Orbital evolution}

We simulated the orbital evolution of the planets by performing \textit{N}-body simulations that calculate the gravitational attraction between all the bodies. Collisions are treated as inelastic mergers, even though the atmospheric mass loss by giant impacts is taken into account in most runs. The physical radii of the planets are calculated assuming a core density of $3 {\rm \,g/cm^3}$ and an envelope density of $1 {\rm \,g/cm^3}$.

\subsection{Long-term simulation of atmospheric loss after formation}

The retention of the accreted hydrogen-rich atmospheres is governed by the thermal evolution of the planets after disk dispersal (1-10\,Myr).
Because the orbital evolution of planets due to influences such as giant impacts continues to occur for 10-100\,Myr after disk dispersal,
we consider the long-term thermal evolution of each planet after the dynamical evolution of the planetary system ceases ($\sim$50\,Myr).
We calculate the atmospheric loss from the planets for 1\,Gyr,
where we assume that the orbital configurations of the planetary systems hardly change during the post-formation phase. As discussed in Section~\ref{sec:PE}, this assumption can be justified according to studies on the orbital stability \citep[e.g.,][]{1996Icar..119..261C}. Note, however, that it has been 
found that the orbital stability can be affected by the long-term change in planetary mass and stellar mass in a very recent study \citep{2020arXiv200301965M}. This effect should be investigated in future work.

\subsubsection{Atmospheric loss by stellar X-ray and extreme ultraviolet irradiations}

Planets in close-in orbits undergo atmospheric loss due to stellar X-ray and extreme ultraviolet (XUV) radiations and the injection of high-energy particles via stellar wind and coronal mass ejections. In this study, we examine the mass loss of a hydrogen-rich atmosphere from a planet as a post-formation process \citep[e.g.,][]{1981Icar...48..150W}.

The mass loss rate from an evaporating planet via energy-limited hydrodynamic escape, $\dot{M}_{\rm esc}$, is given by 
\begin{equation}\label{eq:m_esc}
    \dot{M}_{\rm esc} = \frac{\epsilon F_{\rm XUV} \pi R^3_{\rm XUV}}{G M K_{\rm tide}},
\end{equation}
where $\epsilon$ is the efficiency of heating due to stellar XUV radiation, $F_{\rm XUV}$ is the XUV radiation flux, $G$ is the constant of gravitation, and $K_{\rm tide}$ is the correction factor that accounts for tidal effects in the planetary Roche lobe \citep{2007A&A...472..329E}. The planetary radius, $R_{\rm XUV}$, indicates the radius at which the hydrogen-rich atmospheres become optically thick to XUV photons. As in previous studies \citep[][]{2014A&A...571A..94S,2015SoSyR..49..339I,2018MNRAS.476.5639I}, the heating efficiency, $\epsilon$, is less than 20\% for hydrogen-dominated upper atmospheres. \citet{2012MNRAS.425.2931O} showed that the efficiency for Earth-size planets is low ($\epsilon \sim$ 0.1-0.15).
Therefore, we use a constant value of $\epsilon = 0.1$, even though $\epsilon$ changes with time. 
We define $R_{\rm p}$ as the photosphere, i.e., $R_{\rm p} = R_{\rm bc} + R_{\rm atm}$, where $R_{\rm bc}$ is the radiative--convective boundary and $R_{\rm atm}$ is the photospheric correction given in \citet[][]{2014ApJ...792....1L}.
We integrate the interior structure of planets that undergo atmospheric mass loss and calculate $R_{\rm bc}$ at a given time (see also Section \ref{sec:n-body} for interior models of planets). In this study, we assume that $R_{\rm XUV} \sim R_{\rm p}$.

\subsubsection{Stellar XUV flux}
The temporal evolution of the XUV flux from a Sun-like star remains poorly constrained. We adopt the scaling law of X-ray luminosity for G-dwarfs with ages of $\sim$ 6-740\,Myr given in \citet{2012MNRAS.422.2024J}:
\begin{equation}\label{eq:xuv}
    L_{\rm XUV}(t) =
        \begin{cases}
            L_{\rm sat},\,\,\,\,\,\,\,\,\,\,\,\,\,\,t \leq 700\,{\rm Myr} \\
            L_{\rm sat}\,t^{-1.1},\,\,\,t > 700\,{\rm Myr}, 
        \end{cases}
\end{equation}
where $L_{\rm XUV}$ is the stellar XUV luminosity, $t$ is the stellar age in Gyr, and $L_{\rm sat} = 10^{-3.67} L_\odot$ is the saturated XUV luminosity. As an extreme case, we also adopt a case in which the luminosity was three times higher, $L_{\rm sat} = 10^{-3.19} L_\odot$, than the standard case. We compute the thermal evolution of each planet for up to 1\,Gyr.

\section{Unified simulation of formation and atmospheric evolution}\label{sec:n-body}

\begin{figure*}[ht!]
\plottwo{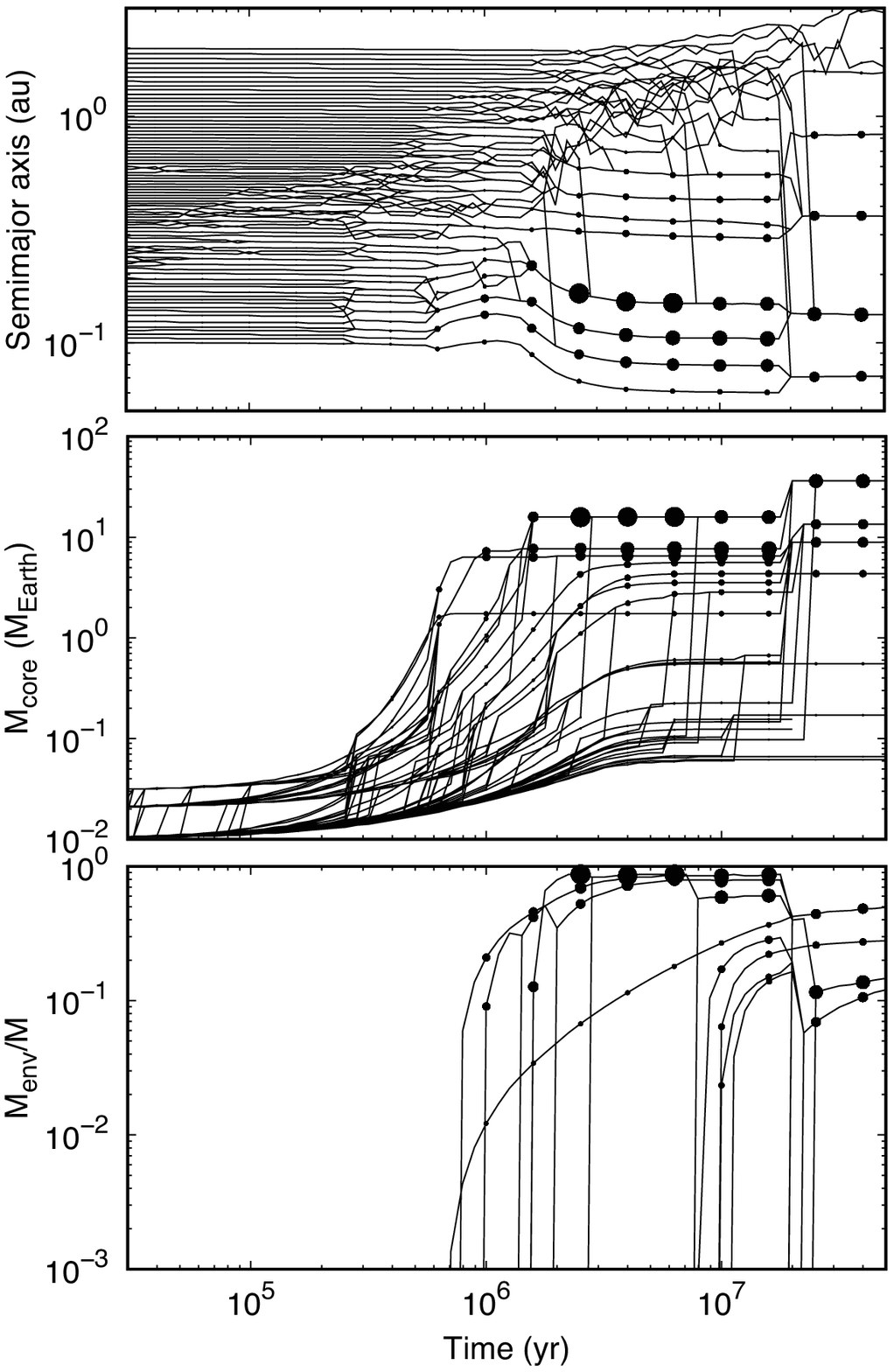}{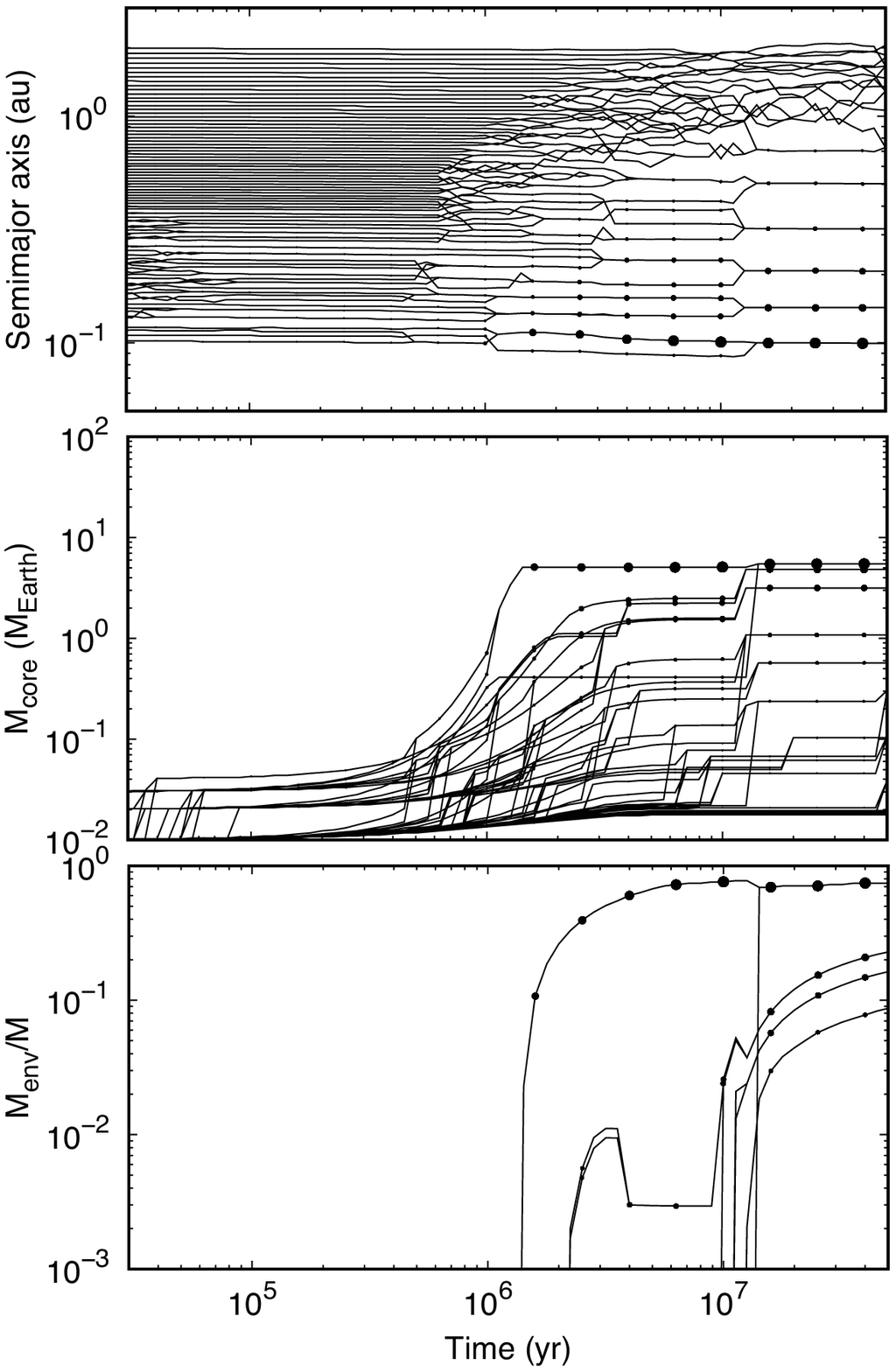}
\caption{Time evolution of the semi-major axis (top), the core mass (middle), and the envelope mass fraction (bottom) of each planet. A typical result for the high pebble flux is shown in the left panel, while one for the low pebble flux is shown in the right panel. In both simulations, the resonant chains undergo orbital instability after gas depletion ($t > 5\,{\rm Myr}$).}
\label{fig:t_a}
\end{figure*}

First, we see some typical outcomes of our unified simulations for pebble-driven planet formation and atmospheric evolution. Figure~\ref{fig:t_a} shows the evolution of the semi-major axis, the core mass, and the envelope mass fraction for our fiducial runs. The left panels show the results for a high pebble flux ($\dot{M}_{\rm pb} = 10^{-4}\,M_\oplus\,{\rm yr}^{-1}$), while the right panels show those for a low pebble flux ($\dot{M}_{\rm pb} = 3.0 \times 10^{-5}\,M_\oplus\,{\rm yr}^{-1}$). 
One of the most remarkable points regarding the semi-major axis evolution is that the planets do not undergo significant migration. As was shown in a previous paper \citep{2018A&A...615A..63O}, type I migration can be significantly suppressed in the close-in region ($r < 1 {\rm \,au}$) due to the decrease in the gas surface density and the change in its slope. Planets actually exhibit slow migration with timescales on the order of 1\,Myr, and as a result most planets are in mean-motion resonances after $t \simeq 1 {\rm \,Myr}$. The chain of resonant planets exhibits an orbital instability after disk gas depletion ($t \simeq 5 {\rm \,Myr}$), leading to giant impacts between the planets. Accordingly, the final orbits are not in mean-motion resonances. Note that, in this paper, we only look at planets that formed inside $r = 1 {\rm \,au}$ because the mass of the planets at $r > 1 \,{\rm au}$ can be affected by the outer boundary ($= 2 {\rm \,au}$) of the initial solid distribution.

Regarding the evolution of the core mass, planetary accretion proceeds from the interior of the disk. In the early stage of planetary formation ($t < 1 {\rm \,Myr}$), even though the planets sometimes undergo collisions, they grow primarily due to pebble accretion. After $t \simeq 1 {\rm \,Myr}$, the planets reach the pebble isolation mass. After that, the cores do not grow substantially because they are in a chain of resonant planets and no collisional events occur. After disk depletion ($t > 5 {\rm \,Myr}$), they grow via giant impacts triggered by the orbital instability of the resonant chain. The typical core mass is approximately $10\,M_\oplus$ for the high pebble flux case, while it is approximately $3\,M_\oplus$ for the low flux case.

Regarding the envelope mass, planets do not accrete massive atmospheres while they undergo pebble accretion ($t < 1 {\rm \,Myr}$). As discussed in the following section, the critical core mass is increased due to heating by pebble accretion. After the planets reach the pebble isolation mass and pebble accretion terminates, the pressure gradient in the envelopes is not strong enough to dominate over the core gravity, leading to a rapid gas accretion onto the core. The envelope mass exponentially increases during runaway gas accretion phase. The envelope accretion is calculated using Eq.~(\ref{eq:mdot_min}); here, the accretion rate onto cores with $M \gtrsim 10\,M_\oplus$ is limited by local or global disk accretion. Even though a fraction of the atmosphere is lost during giant impacts ($t > 10 {\rm \,Myr}$), the envelope mass fraction is large in the final state. Note that the planets accrete a small amount of H$_2$/He atmosphere from the small-mass remnant disk in the very late stage $(t > 10 {\rm \,Myr})$. We expect that such late-stage atmospheric accretion may not occur when photo-evaporation clears the inner disk after disk dispersal $(t > 5 {\rm \,Myr})$. We will discuss the effect of disk photoevaporation, e.g., the disk evolution model developed in \citet{2020MNRAS.492.3849K}, on the atmospheric growth of close-in super-Earths in our next paper.

\begin{figure*}[ht!]
\plottwo{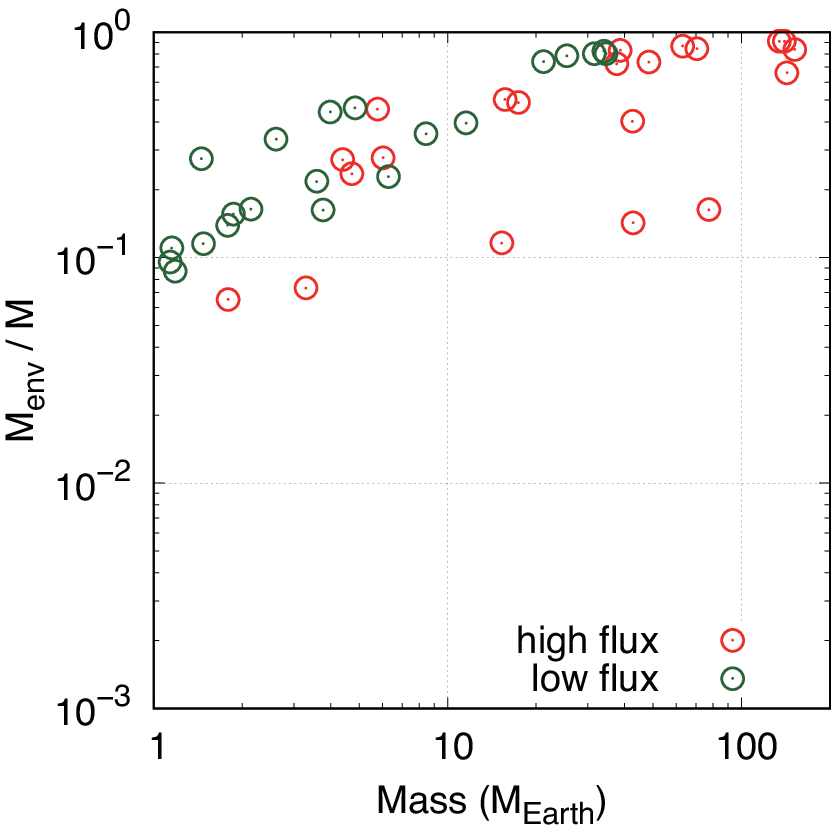}{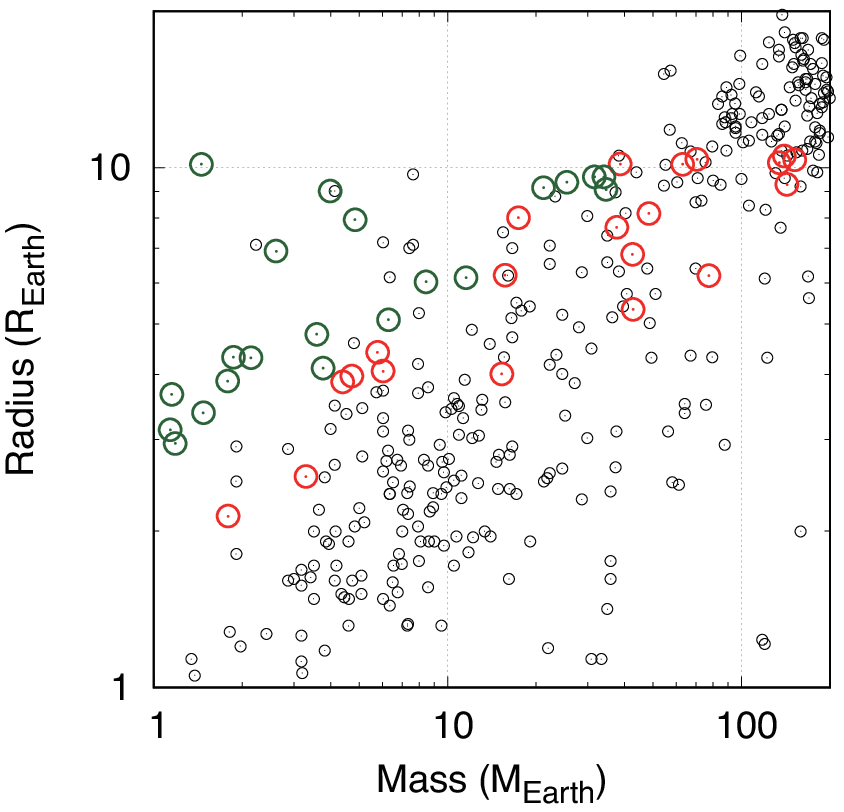}
\caption{Final envelope masses (left) and mass-radius relationship (right) of the planets after 50\,{\rm Myr}. The red and green circles represent the results for the high pebble flux and low pebble flux cases, respectively. Five simulation runs were performed for each model. The small black circles on the mass-radius diagram indicate confirmed exoplanets. The simulated planets accrete large atmospheres, which is inconsistent with observations.}
\label{fig:m_n}
\end{figure*}

Figure~\ref{fig:m_n} summarizes the results of five simulation runs for each case with different pebble fluxes. Each panel shows the envelope mass fraction and the planetary radius of the simulated planet after 50\,{\rm Myr}. The planetary radius is calculated such that the interior structure of the planet is integrated using two equations of state (EoSs), the SCvH EoS \citep{1995ApJS...99..713S} for a hydrogen-rich atmosphere and the Vinet EoS for a rocky core whose thermodynamic properties, such as the bulk moduli, are taken from \citet{2009JGRB..114.1203M}.
We find that the envelope mass fraction is approximately 10-90\%, which is inconsistent with estimates of the envelope mass fractions of observed transiting super-Earths (0.1-10\% by mass on average)  \citep[e.g.][]{2014ApJ...792....1L}. In fact, the mass-radius diagram shows that super-Earths of $\lesssim 10\,M_{\oplus}$ in our simulations are puffed up, compared to observed ones. We also find that the final planetary mass depends on the pebble flux. An interesting result is that only a factor of 3 difference in the pebble flux results in a factor of 10 mass difference, which was also found in \cite{2019A&A...627A..83L}.

\subsection{Suppression of the runaway gas accretion by pebble accretion}\label{sec:suppress}
We then take a closer look at three mechanisms for avoiding accretion and the retention of massive atmospheres shown in Section~\ref{sec:intro}. First, we focus on whether the accretion of massive atmospheres could be avoided due to heating by pebble accretion. As we saw in Figures~\ref{fig:t_a} and \ref{fig:m_n}, the accretion of massive atmospheres cannot be avoided.

The critical core mass for rapid gas accretion increases with the pebble accretion rate.
According to Equation~(\ref{eq:mcrit}), the critical core mass exceeds $10\,M_\oplus$ for $\dot{M}_{\rm pb} > 3 \times 10^{-7}\,M_\oplus\,{\rm yr}^{-1}$. This means that, even when the pebble accretion rate is relatively small, the accretion of massive atmospheres can be delayed by pebble heating. However, planets reach the pebble isolation mass before the disk dispersal. After the planets reach the pebble isolation mass, pebble heating does not exist and the critical core mass decreases, leading to the accumulation of massive atmospheres.

To avoid runaway gas accretion, it is necessary that the time to reach the pebble isolation mass be longer than the disk lifetime. In addition, planets should grow to super-Earth masses ($\simeq 5\,M_\oplus$) and the growth timescale should not be too long. Therefore, to form super-Earths with small atmospheres, it is necessary that the time to reach the pebble isolation mass be comparable to the disk lifetime. It is very difficult to satisfy this condition without tuning the parameters. The pebble accretion rate increases with the planetary mass (see Section~\ref{sec:pebble}). 
Planets that reach 1 Earth mass grow to the pebble isolation mass in a short time (see also Figure\,4 in \citealt{2017AREPS..45..359J}), leading to a shutoff of the pebble heating. Therefore, a fine-tuning of the parameters is needed to satisfy the above condition\footnote{We performed additional simulations with the pebble flux further reduced by a factor of three and found that the cores did not grow to super-Earths.}. This condition may be satisfied if the pebble accretion rate somehow decreases after the planets reach 1 Earth mass.

\subsection{Atmospheric loss due to giant impacts}\label{sec:atm-loss}

\begin{figure}[ht!]
\plotone{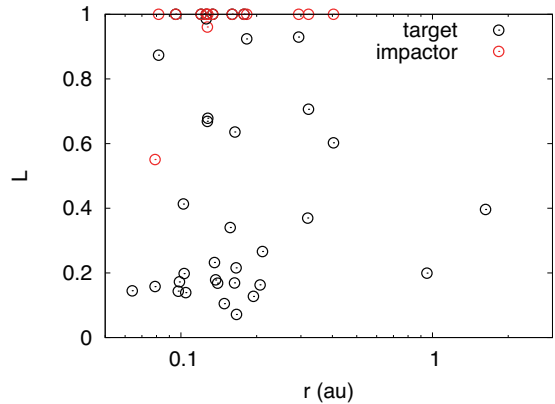}
\caption{Atmospheric loss fraction during each impact event for five simulations with the high pebble flux. Planets typically lose approximately 20\% of accreted H$_2$/He atmosphere in a single collision.}
\label{fig:r_L}
\end{figure}

Next, we examine the effect of atmospheric loss during giant impacts. We analyze each impact event after the planets start to accrete H$_2$/He atmospheres and find that the typical impact velocity is $1-2\,v_{\rm esc}$, where $v_{\rm esc}$ is the mutual escape velocity. Figure~\ref{fig:r_L} shows the atmospheric loss fraction ($L = M_{\rm loss}/M_{\rm env}$) as a function of the radial distance for five simulation runs with the high pebble flux, where $M_{\rm loss}$ is the atmospheric mass that is lost by one impact and $M_{\rm env}$ is their atmospheric mass before the impact. The collision data are plotted only when the target or the impactor have an atmosphere of more than $0.1\,M_\oplus$. We see in Figure~\ref{fig:r_L} that the atmosphere of the impactor is significantly eroded, which can be explained by the dependence of the impact energy on the mass ratio in Eq.~(\ref{eq:Qs}). 
Regarding the atmospheric loss from the targets, the mass loss fraction ranges from a few percent to approximately 90\% and the typical mass loss fraction is approximately 20\%. As seen in Figure~\ref{fig:t_a}, planets undergo only one or two collisional events after they acquire atmospheres\footnote{This number of giant impact events is consistent with the results of previous \textit{N}-body simulations \citep[e.g.,][]{2017MNRAS.470.1750I, 2018A&A...615A..63O}}. Therefore, we conclude that if planets accrete massive atmospheres ($\gtrsim 50 \%$), even atmospheric loss due to giant impacts cannot make them super-Earths with a small amount of atmosphere ($\lesssim 10 \%$). Although not shown here, we confirmed that the amount of atmospheric loss is larger for collisions with smaller impact parameters such as head-on collisions. 

\begin{figure}[ht!]
\plotone{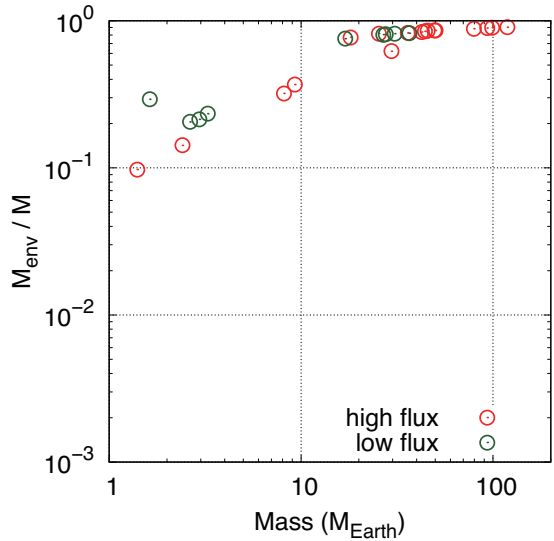}
\caption{Same as in Figure~\ref{fig:m_n} but without the effects of atmospheric loss during giant impacts.}
\label{fig:m_menv}
\end{figure}

Figure~\ref{fig:m_menv} shows the final envelope mass fraction after 50\,Myr for a total of 10 runs of additional simulations (five runs for the high pebble flux and five runs for the low pebble flux). In these simulations, the atmospheric loss during the giant impacts is not considered for comparison. The final envelope mass fraction is smaller for cases in which the impact erosion is included (Figure~\ref{fig:m_n}); however, there is no significant difference between the two sets of simulations. This confirms that the mass loss during giant impacts cannot significantly reduce the amount of the atmosphere.

\subsection{Orbital and physical properties of super-Earths}\label{sec:property}
In the previous section, we demonstrated that the envelope mass fraction of the observed super-Earths cannot be reproduced by the results of our standard simulations. In this section, we compare orbital properties of observed planets with those of simulated planets.

\begin{figure}[ht!]
\plotone{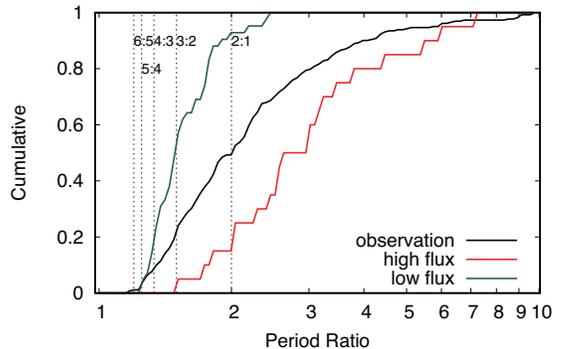}
\caption{Cumulative period-ratio distributions of the planetary systems. The red and green lines represent summaries of the five simulation runs for the high pebble flux and low pebble flux cases, respectively. The black line indicates the distribution for the observed super-Earths.}
\label{fig:p_n}
\end{figure}

Figure~\ref{fig:p_n} compares the cumulative distribution of the period ratio for adjacent planets in the same way as in \cite{2018A&A...615A..63O}.
We see that the period ratio is smaller for the case in which the pebble flux is smaller. Because the final planetary mass increases with increasing pebble flux, the period ratio is smaller for smaller planets. Note that the orbital separation is between $10\,r_{\rm H}$ and $40\,r_{\rm H}$ (typically $20\,r_{\rm H}$), where $r_{\rm H}$ is the mutual Hill radius, irrespective of the planetary mass. This explains why the period ratio is smaller for smaller planets because the Hill radius depends on the planetary mass such that $r_{\rm H} \propto M^{1/3}$. 
As shown in \cite{2018A&A...615A..63O}, by blending the two cases with different pebble fluxes, the period ratio distribution of the observed super-Earths can be better reproduced. 
Note that in this paper, a relatively high alpha viscosity ($\alpha_{r,\phi} = 8 \times 10^{-3}$) was used and the type I migration is significantly suppressed. When we adopt a smaller value of viscosity, it was shown that planets are more prone to migration due to differences in the disk profile and the effect of desaturation of the corotation torque \citep{2018A&A...615A..63O}.
Note also that if the gas surface density behaves like a power-law distribution as derived by \citet{2016ApJ...821...80B}, planets undergo inward migration.
Although not shown here, the mass distribution is also matched by the observed mass distribution by blending cases with different pebble fluxes (see also \citealt{2018A&A...615A..63O}). 

According to \cite{2017MNRAS.470.1750I}, 90-95\% of the system should undergo the late orbital instability after disk gas depletion in order to match the observed period ratio distribution. We confirm that most planetary systems in our simulations undergo the late orbital instability. The late instability is observed in all five simulations for high pebble flux, while the instability is not seen in two simulations out of five for low pebble flux. \citet{2018AJ....155...48W} pointed out that \textit{Kepler} multi-planet systems have remarkable properties; they are similar in size and regularly spaced \citep[e.g.][]{2011ApJS..197....8L}. The orbital period ratios are smaller in systems with smaller planets. All of these properties can be explained by results of our simulations without tuning of parameters.

Regarding orbital migration, previous studies \citep{2015A&A...578A..36O, 2017A&A...607A..67M} have shown that compact systems in mean-motion resonances are produced due to rapid type I migration, which is inconsistent with the observations. In contrast, as we saw in Figure~\ref{fig:t_a}, our planets do not undergo significant type I migration and, as a result, their orbital properties are consistent with the observed distributions. Even though this was already shown in \cite{2018A&A...615A..63O}, we confirm that the inclusion of pebble accretion does not alter this trend. 

Finally, we comment here on the composition of the planetary core. According to \cite{2019arXiv190208772I}, planetary cores that grow outside the snow line move into the close-in region to form close-in super-Earths. As a result, these cores primarily consist of ice, which may be inconsistent with the inferred core composition of super-Earths \citep[e.g.,][]{2017ApJ...847...29O}. In our simulations, planetary cores do not undergo significant migration and close-in super-Earths presumably consist of refractory materials. This is consistent with estimates of the compositions of some super-Earths that may consist of refractory materials \citep[e.g.,][]{2019MNRAS.484..712D}.

\section{Long-term simulation of atmospheric loss after formation}\label{sec:PE}

Next, we compute long-term evolution of planets having H$_2$/He atmospheres for $\sim 1$\,Gyr and examine whether massive atmospheres can be evaporated by stellar X-ray and EUV irradiations. In these extended simulations, we assume that the planetary orbit is fixed. This assumption can be justified because the orbital configuration would not change significantly during this stage. According to studies on the orbital stability \citep[e.g.,][]{1996Icar..119..261C}, the orbital stable time of systems with typical orbital separations of $20 \,r_{\rm H}$ (Section~\ref{sec:property}) would be very long ($> 1 {\rm \,Gyr}$). Nevertheless, the effect of long-term change in the planetary mass \citep{2020arXiv200301965M} should be investigated in future work.

\begin{figure*}[ht!]
\plottwo{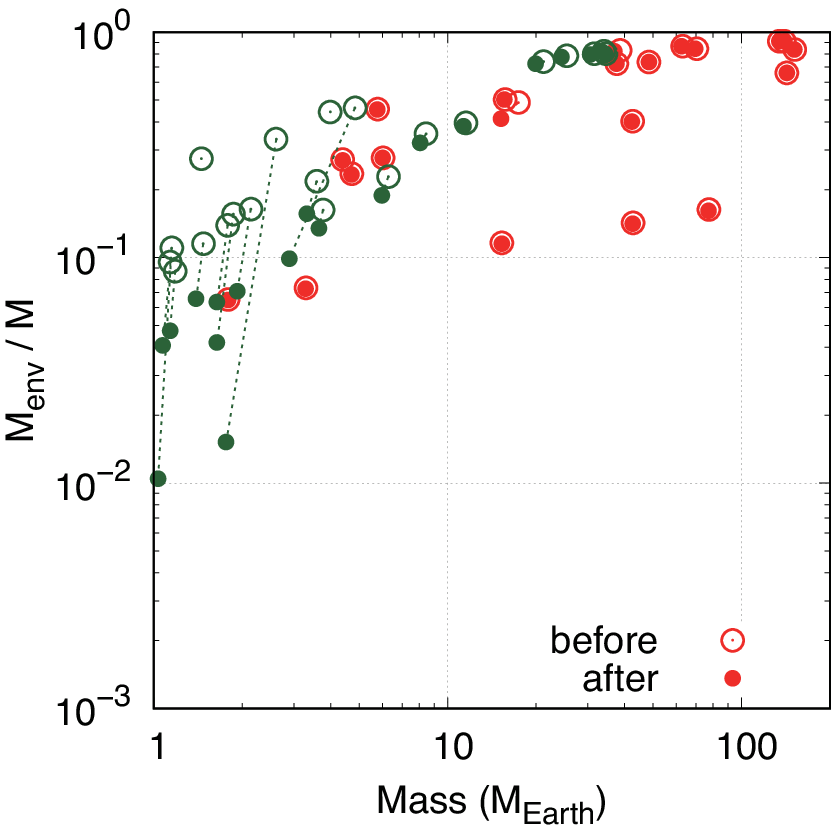}{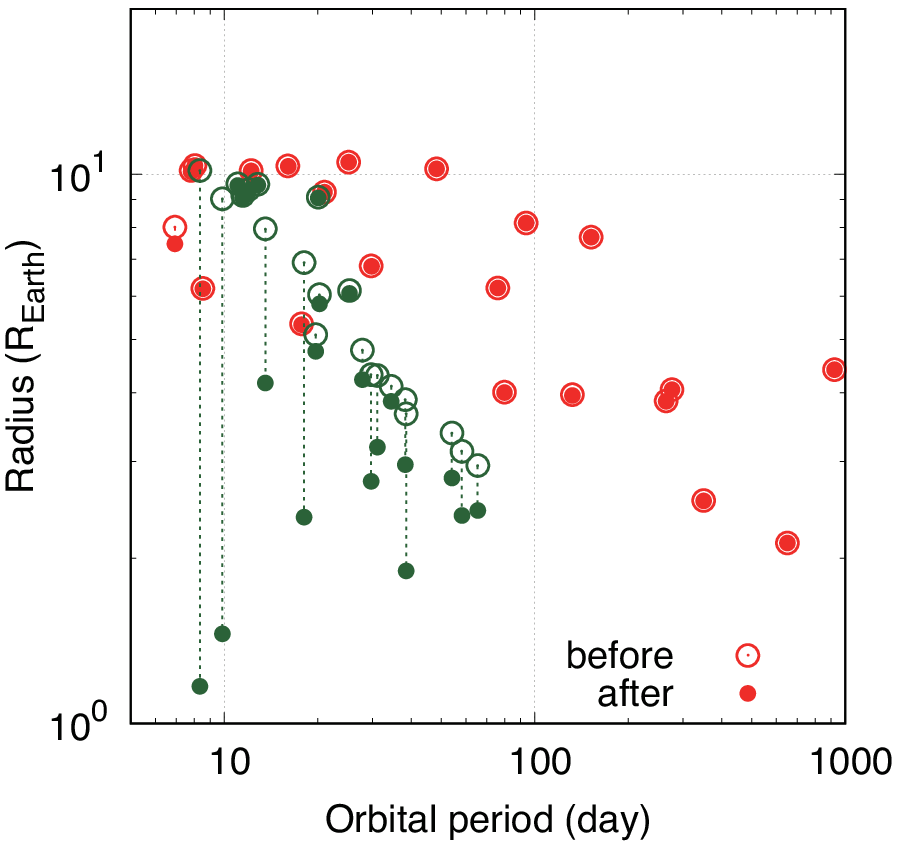}
\caption{Envelope mass fractions (left) and radius-period distribution (right) of evaporating planets right after the formation stage (50\,Myr: open circles) and after 1\,Gyr (filled circles). The atmospheric evolution track of each planet is shown by a dotted line. The red and green circles represent the results for the high pebble flux and low pebble flux cases, respectively.}
\label{fig:PE}
\end{figure*}

The left panel of Figure~\ref{fig:PE} shows the envelope mass fraction at the end of the formation stage $t = 50 {\rm \,Myr}$ and after the long-term evolution $t = 1 {\rm \,Gyr}$. The atmospheric mass loss rate depends on orbital and physical properties of planets, $\dot{M}_{\rm esc} \propto R^3_{\rm p}/(M\,r^{2})$ (see Eq.~(\ref{eq:m_esc})). 
We find that, even though smaller planets in close-in orbits ($M \lesssim 10\,M_\oplus$, $P \lesssim 50$\,days) lose 30-40\% of the accreted H$_2$/He envelope for 1\,Gyr, the atmospheric loss from planets is typically $\lesssim 1 \,M_\oplus$. A large fraction of the atmosphere remains at the end of the simulation. Planets that formed in the high pebble flux are large enough to survive against photo-evaporation. Typical evaporated fractions for 1\,Gyr are approximately 0.1\% for the high pebble flux cases and a few percent for the low pebble flux cases. In addition, we cannot perfectly reproduce the observationally inferred radial valley of $R \simeq 1.5-2.0 \,R_\oplus$ in the radius-period diagram \citep[e.g.,][]{2017AJ....154..109F, 2017ApJ...847...29O}. In the right panel of Figure~\ref{fig:PE}, planets tend to not exist in the region of $R\simeq 1.5-6\,R_\oplus$, which is wider than the inferred radius valley. This is primarily because the thickness of the atmosphere is too large at the end of the formation stage.

We also performed long-term simulations of the photo-evaporation assuming a three times larger XUV luminosity than in Eq.\,(\ref{eq:xuv}), which represents a case of high luminosity for G stars \citep{2012MNRAS.422.2024J}. 
In this case, even though the mass loss rate from planets increases under intense XUV irradiations, it is unlikely to evaporate massive H$_2$/He atmospheres by more than $\simeq 30$\,wt\%. The typical envelope mass fractions at the end of the simulations are $0.1-90\%$. As seen in Figure~\ref{fig:PE}, the envelope mass fraction of planets with $M \gtrsim 10\,M_\oplus$ remains approximately unchanged.

\section{Limit on gas accretion due to disk accretion}\label{sec:disk_limit}

So far, we find that, when using standard models of planetary formation and atmospheric evolution, super-Earths with large H$_2$/He atmospheres ($\gtrsim 10$\,wt\%) form, which is inconsistent with observations. The suppression or delay of rapid gas accretion is necessary for pebble accretion model to explain the mass-radius relationships of transiting super-Earths. In this section, we consider a case in which the atmospheric accretion onto the core is limited by radial mass accretion \citep{2018ApJ...867..127O}.
The underlying assumption is that a rapid gas flow near the disk surface driven by the wind torque (wind-driven accretion) passes through the planets and do not accrete onto them\footnote{Effects of gas flows on the atmospheric accretion should be investigated by 3D hydrodynamic simulations.}.
As stated in Section~\ref{sec:model_env}, we assume that the disk accretion expressed in Eq.~(\ref{eq:mdot_disk}), which contributes to the atmospheric accretion, is regulated by the viscous accretion rate ($\dot{M}_{\rm disk} = \dot{M}_{\rm visc}$).

\subsection{Unified simulation of formation and atmospheric evolution}

\begin{figure}[ht!]
\plotone{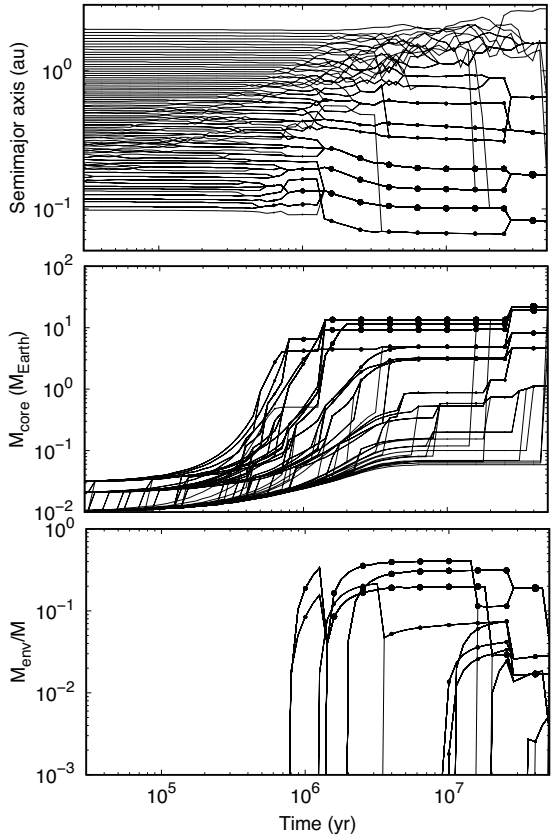}
\caption{
Same as in Figure~\ref{fig:t_a} but considering a limit on the gas supply to the planet, $\dot{M}_{\rm disk} = \dot{M}_{\rm visc}$. A typical run with the high pebble flux is shown.}
\label{fig:t_a_513}
\end{figure}

Figure~\ref{fig:t_a_513} shows a typical result of an \textit{N}-body simulation for the high pebble flux case but with no gas supply from wind-driven accretion to a planet. A major difference with respect to Figure~\ref{fig:t_a} is that the final envelope mass fraction is small. Other characteristics such as the orbital evolution and the core mass are the same as in Figure~\ref{fig:t_a}.

\begin{figure}[ht!]
\plotone{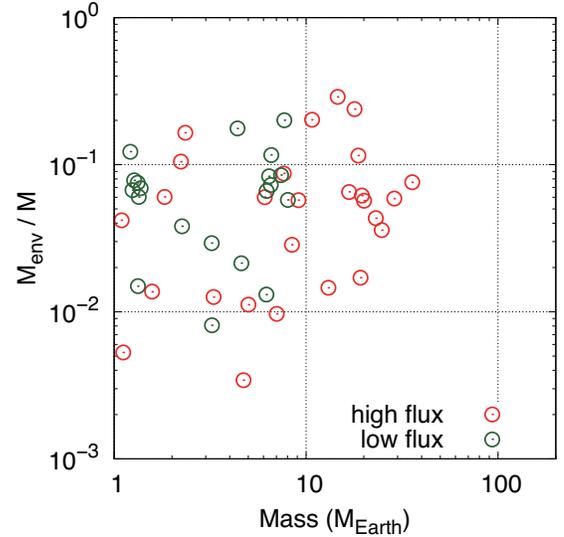}
\caption{Same as in Figure~\ref{fig:m_n} (left) but considering $\dot{M}_{\rm disk} = \dot{M}_{\rm visc}$.}
\label{fig:m_menv_410}
\end{figure}

Figure~\ref{fig:m_menv_410} shows the envelope mass fraction after 50\,My for five runs of each case (high pebble flux and low pebble flux). As demonstrated in \cite{2018ApJ...867..127O}, the final envelope mass fraction is typically less than approximately 10\%. In addition, we find that the period ratio distribution matches the observations when blending the results of different pebble fluxes. 

\subsection{Long-term simulation of atmospheric loss after formation}

\begin{figure*}[ht!]
\plottwo{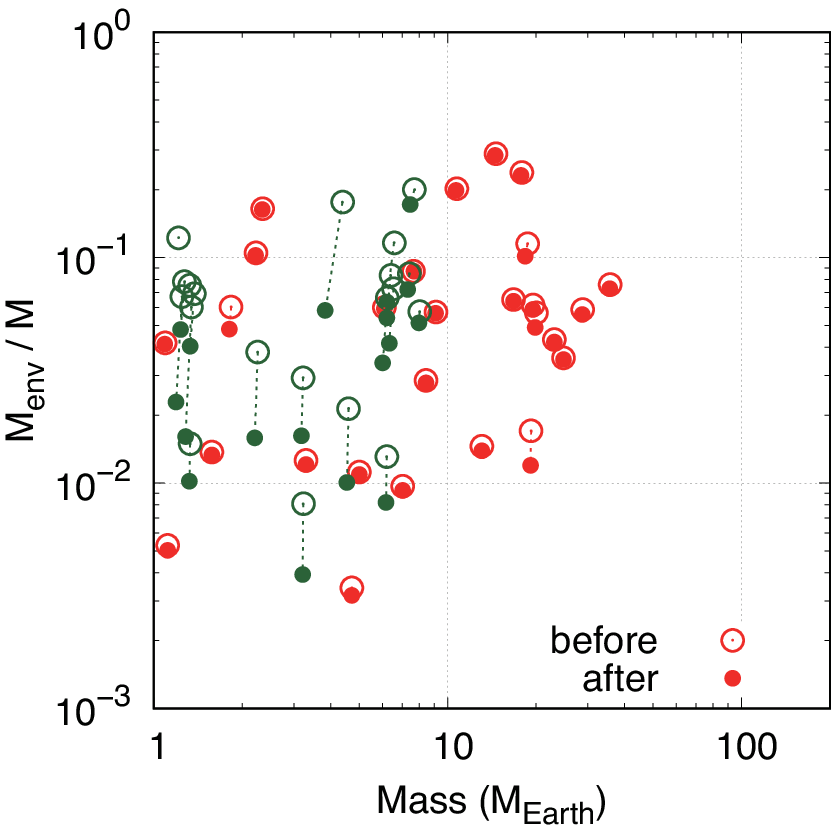}{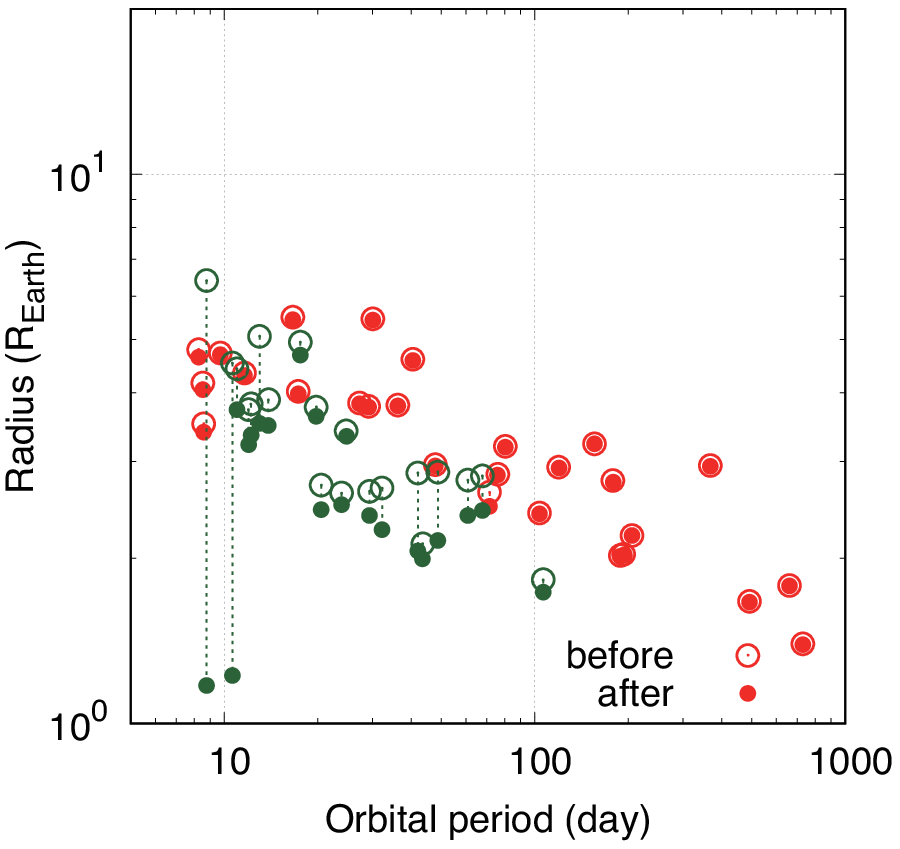}
\caption{Same as in Figure~\ref{fig:PE} but considering $\dot{M}_{\rm disk} = \dot{M}_{\rm visc}$.}
\label{fig:PE_410}
\end{figure*}

We then examine the results of subsequent simulations of the atmospheric loss for 1\,Gyr. Figure~\ref{fig:PE_410} shows a summary of the evolution of the envelope mass fraction and the planetary radius due to photo-evaporation. The evaporated envelope mass is similar to that shown in Figure~\ref{fig:PE}, which is approximately 0.1-10\% for the standard XUV luminosity. As shown in Figure~\ref{fig:m_menv_410}, the envelope mass fraction is less than approximately 10\% after formation. Therefore, large fractions of the accreted H$_2$/He envelopes can be eroded for some planets. Regarding the radius valley, the radius valley seen in Figure~\ref{fig:PE_410} ($R \simeq 1.5-3\,R_\oplus$) is narrower than that shown in Figure~\ref{fig:PE} and more consistent with the observationally inferred region ($R \simeq 1.5-2.0\,R_\oplus$). In addition, it appears that low pebble flux models are more favorable for the observed radius distribution of exoplanets. More statistical arguments are needed for a further discussion of this result, which is beyond the scope of this paper. 

In summary, we find that, if super-Earths do not accumulate massive atmospheres during the formation stage, the envelope mass fraction and the radius valley can be naturally reproduced. This is consistent with the assumption in \cite{2017ApJ...847...29O}, in which the origin of the radius valley is explained by the evaporation of atmospheres from cores that are less than 10\% of the atmosphere. \cite{2019MNRAS.488L..12N} also pointed out that the envelope accretion onto the cores should be suppressed by approximately an order of magnitude to match the planetary mass function inferred by the ALMA observations.

\section{Discussion and Conclusions}\label{sec:conc}

We investigated planetary formation in the close-in region by performing unified \textit{N}-body simulations. In this paper, we focused on the origin of the observed super-Earths and their H$_2$/He atmospheres. Our main findings were as follows.
\begin{itemize}
\item As seen in a previous study \citep{2018A&A...615A..63O}, the observed orbital characteristics are well matched by the results of our simulations. Contrary to some previous studies \citep[][]{2015A&A...578A..36O, 2017A&A...607A..67M}, type I migration is significantly suppressed in a disk evolving with magnetically driven disk winds \citep{2016A&A...596A..74S}. As a result, super-Earths do not form in compact configurations near the inner edge of the disk. Instead, super-Earths undergo slow type I migration and are temporarily captured in a chain of mean-motion resonances, which exhibit late orbital instability after disk depletion. Super-Earths undergo giant impacts, and resonant configurations are lost at the end of the evolution. Owing to the late orbital instability, the observed period ratio distribution is well reproduced. In addition, other orbital properties such as similar sizes and regular spacing \citep[e.g.,][]{2018AJ....155...48W} are naturally explained.
\item Our investigation of the atmospheric evolution of super-Earths stressed the issue of the accumulation of massive atmospheres by super-Earths. At the end of our simulations, the super-Earths possess approximately 1--90\% H$_2$/He atmospheres by mass, which is inconsistent with estimates of the envelope mass fraction for observed super-Earths ($\sim 0.1-10\%$). We examined three possible mechanisms to reproduce these observations. The first possibility is heating of the envelope via pebble accretion. We developed a formula for the critical core mass including pebble heating. We found that pebble heating can efficiently suppress gas accretion onto the cores. However, pebble accretion is quenched when planets reach the pebble isolation mass. After that, super-Earths can accrete massive H$_2$/He atmospheres before the disk gas disappears. The second possibility is atmospheric loss during giant impacts. We found that the typical envelope mass fraction that is lost during giant impacts is approximately 20\%; however, super-Earths undergo only one or two giant impact events, which is not enough to significantly decrease a massive atmosphere. The third possibility is photo-evaporation of the atmospheres due to stellar irradiation. We demonstrated that a massive atmosphere ($>30\%$) cannot be lost due to photo-evaporation even in the extreme case of high XUV luminosity. Therefore, in the standard setting of current planetary formation theory, super-Earths with massive H$_2$/He atmospheres remain. In other words, gaseous planets are easy to form in the close-in region. This is contrary to the observational results.
\item Consequently, this study suggested that there are mechanisms that operate during the formation stage to keep the envelope mass fraction smaller than approximately 10\% by mass. Adopting one method, in which the atmospheric accretion is limited by disk accretion \citep{2018ApJ...867..127O}, we reran the simulations. We found that, when the atmospheric accretion is limited, several observed properties of super-Earth atmospheres (e.g., the envelope mass fraction) can be reproduced.
\end{itemize}

In this paper, we adopted a relatively high value of the alpha viscosity $\alpha_{r,\phi} = 8 \times 10^{-3}$. This is because in the previous study, the observed orbital properties of super-Earths were reproduced with this viscosity \citep{2018A&A...615A..63O}. If a smaller value of the turbulent viscosity is used (e.g., $\alpha_{r,\phi} = 8 \times 10^{-5}$), planets are more prone to migration and the final orbits of super-Earths are in mean-motion resonances more often \citep{2018A&A...615A..63O}. The pebble accretion rate is higher for smaller alpha viscosity because the pebble scale height becomes smaller (Eq.~(\ref{eq:hpb})). However, the qualitative outcomes of atmospheric evolution remain unchanged. That is, super-Earths accrete massive atmospheres after they reach the pebble isolation mass. We note that although the turbulent strength can become small in the outer region at a few tens of au \citep[e.g.][]{2017ApJ...843..150F,2018ApJ...856..117F}, the value of the alpha viscosity can be relatively high in the close-in region, especially at the late stage of disk evolution \citep[e.g.][]{1996ApJ...462..725G,2004ApJ...603..213C,2015ApJ...811..156D,2019ApJ...871...10U}.

We used the disk evolution model developed by \cite{2016A&A...596A..74S}. As stated in Section~\ref{sec:disk}, \cite{2016ApJ...821...80B} derived a different disk evolution model like a power-law distribution. The difference can be attributed to adopted prescriptions for the mass loss due to disk winds and the evolution of the vertical magnetic field (see Section~4.4 of \citealt{2016A&A...596A..74S}). In \cite{2016ApJ...821...80B}, the gas surface density behaves like a power-law distribution ($\Sigma_{\rm g} \propto r^{-(1-1.5)}$). In such a disk, super-Earths undergo inward migration, and concentrate towards the disk inner edge \citep[e.g.][]{2019A&A...627A..83L}, or fall onto the star \citep[][]{2018A&A...615A..63O}. It would be interesting to investigate the atmospheric evolution of super-Earths in such a disk; it is likely that super-Earth cores accrete massive atmospheres, as seen in our simulations.

In future studies, we need to perform simulations considering different mechanisms that may limit the atmospheric accretion. It would be interesting to see whether the possibilities raised in Section~\ref{sec:intro} actually help to form super-Earths with small atmospheres. For example, the envelope can be polluted by accreting pebbles \citep[e.g.][]{2019ApJ...871..127V}. The polluted envelope layer above the core would delays the envelope cooling \citep{2011MNRAS.416.1419H,2015A&A...576A.114V}.
Recently, \cite{2020A&A...634A..15B} derived an analytical expression for the critical core mass for a polluted envelope in the pebble accretion scenario. Since silicate pebbles can grow via collisions in the envelope, they should settle down in a deep interior and then evaporate. The metal pollution by accreted pebbles in the envelope would affects the thermal state of a planet, namely, atmospheric contraction. The effect of pebble-driven pollution on the critical core mass will be discussed in future work.
As a different mechanism, the atmospheric loss during giant impacts can be updated. It is likely that the atmospheric loss fraction increases when we consider the presence of water \citep{2005Natur.433..842G} or the thermal components of H$_2$/He atmospheres \citep{2019MNRAS.485.4454B}.

\acknowledgments

We thank Simon Lock and Sarah Stewart for help with the model of atmospheric loss. We also thank the anonymous referee for a thorough review.
This work was supported by JSPS KAKENHI Grant Number 18K13608 and 19H05087.
Numerical computations were in part carried out on PC cluster at Center for Computational Astrophysics of the National Astronomical Observatory of Japan.

\appendix
\section{Relation between critical core mass and pebble accretion rate}\label{app:mcrit}

A critical core mass means how massive core can maintain hydrostatic equilibrium of the planet's interior.
Once a planetary core reaches a critical core mass through the accretion of solids such as pebbles and planetesimals,
it goes into runaway gas accretion.
Runaway gas accretion is triggered by gravitational contraction of the envelope onto a core.
The envelope heating by the accretion of solids increases the local pressure gradient that supports the core gravity, leading to the delay of runaway gas accretion. As a result, a critical core mass is positively correlated with the accretion rate of solids.

We determine a critical core mass as a function of pebble accretion rate in the following way.
Given that a planetary interior is in hydrostatic and thermodynamic equilibrium,
we calculate how massive envelope can exist above the surface of a core.  
The interior structure of a planet is described by fundamental equations that govern stellar structure and evolution \citep[see][]{1990sse..book.....K}.
Heat transfer in the envelope of a planet is controlled by either convection or radiation.
We assume that the internal luminosity $L$ results from the accretion of pebbles, which is given by
$L = G M_{\rm core} \dot{M}_{\rm peb}/R_{\rm core}$,
where $G$ is the gravitational constant, $M_{\rm core}$ is the core mass, $R_{\rm core}$ is the core radius,
and $\dot{M}_{\rm peb}$ is the accretion rate of pebbles.
While increasing $M_{\rm core}$, we repeatedly simulate the hydrostatic structure of a planet that grows at a given $\dot{M}_{\rm peb}$.
We find the local maximum of a core mass, which corresponds to a critical core mass for $\dot{M}_{\rm peb}$, under the condition.

\begin{figure}[ht!]
\includegraphics[width=0.5\textwidth]{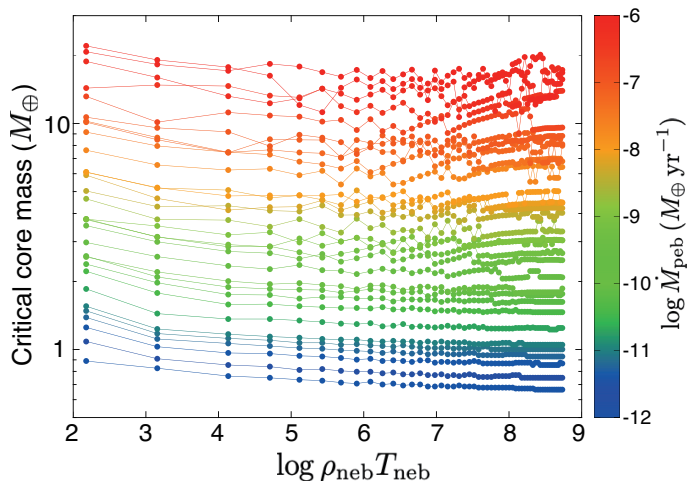}
\caption{Relation between a critical core mass and a pebble accretion rate ($10^{-12}$--$10^{-6}\,M_\oplus\,{\rm yr}^{-1}$ from the bottom to the top). The horizontal axis corresponds to the choice of the density ($\rho_{\rm gas}$) and temperature ($T_{\rm gas}$) of a disk gas as outer boundary conditions.}
\label{fig:mcrit}
\end{figure}

Figure\,\ref{fig:mcrit} demonstrates the relation between a critical core mass and a pebble accretion rate. 
We use opacity tables of ISM-like dust grains given by \citet{2003A&A...410..611S} and those of gas given in \citet{1994ApJ...437..879A}.
A rapid growth of small grains through collisions may lead to the depletion of grains in the envelope. 
We consider that grain opacities can be reduced to 1\% of the ISM values \citep[e.g.][]{2008Icar..194..368M}.
Although pebbles may dissociate and sublimate in the envelope,
we do not consider envelope pollution by pebbles in this study, namely, changes in chemical compositions and opacities in the envelope.
Note that high $\dot{M}_{\rm peb}$ yields a large critical core mass as mentioned above.
A critical core mass is as small as $1\,M_\oplus$ in low $\dot{M}_{\rm peb}$ cases ($\lesssim 10^{-10}\,M_\oplus\,{\rm yr}^{-1}$) because the outermost isothermal layer extends deep in the envelope. 
Unless the planetary interior is wholly convective, a critical core mass is insensitive to the choice of outer boundary conditions, i.e., the density and temperature of a disk gas, as seen in Figure\,\ref{fig:mcrit} \citep[see also][]{1982P&SS...30..755S}. Thus, we find a fitting formula of a critical core mass as a function of pebble accretion rate (see Eq.\,\ref{eq:mcrit}).
In our {\it N}-body simulations, if a planetary core that grows at a given $\dot{M}_{\rm peb}$ exceeds the critical core mass given by Eq.\,\ref{eq:mcrit}, it starts runaway gas accretion.

\bibliography{reference}
\bibliographystyle{aasjournal}



\end{document}